\documentclass[showpacs,12pt,aps,prl]{revtex4}
\usepackage{graphicx}
\begin{document}
\bibliographystyle{apsrev}

\title{ Memory Functions of the Additive Markov Chains:\\
Applications to Complex Dynamic Systems}

\author{S. S. Melnyk, O. V. Usatenko, and V. A. Yampol'skii
\footnote[1]{E-mail: yam@ire.kharkov.ua} }
\affiliation{A. Ya. Usikov Institute for Radiophysics and Electronics \\
Ukrainian Academy of Science, 12 Proskura Street, 61085 Kharkov,
Ukraine}

\begin{abstract}
A new approach to describing correlation properties of complex
dynamic systems with long-range memory based on a concept of
\emph{additive} Markov chains (Phys. Rev. E 68, 061107 (2003)) is
developed. An equation connecting the memory and correlation
function of the system under study is presented. This equation
allows reconstructing a memory function using a correlation
function of the system. Effectiveness and robustness of the
proposed method is demonstrated by simple model examples. Memory
functions of concrete coarse-grained literary texts are found and
their universal power-law behavior at long distances is revealed.
\end{abstract}

\maketitle

The problem of long-range correlated dynamic systems (LRCS) has
been under study for a long time in many areas of contemporary
physics~\cite{bul,sok}, biology~\cite{vossDNA,stan},
economics~\cite{stan,mant}, etc.~\cite{stan,czir}. An important
example of complex LRCS are naturally written
texts~\cite{schen,kokol,uyakm}. The efficient method for
investigating long-range correlations in such systems consists in
the decomposition of the space of states into a finite number of
parts labelled by definite symbols, which are naturally ordered
according to the dynamics of the system. The most frequently used
method of the decomposition is based on the introduction of two
parts of the phase space. In other words, the approach assumes
mapping two kinds of states onto two symbols, say 0 and 1. Thus,
the problem is reduced to investigating the statistical properties
of binary sequences.

It might be thought that the coarse graining could result in
losing, at least, the short-range memory in the sequence. The
authors of Ref.~\cite{nar1,nar2} considered the Markov sequences
with a \emph{many-valued alphabet}. They demonstrated that the
mapping of a given sequence into a small-alphabet sequence does
not necessarily imply that the long-range correlations presented
in the initial text would be preserved. Moreover, in general, the
coarse-graining procedure could lead to spurious long-range
correlations. However, as was shown in Ref.~\cite{uyakm},
coarse-graining \emph{does not destroy} the existing correlations
in many real symbolic systems. The statistical properties of
coarse-grained texts depend, but not significantly, on the kind of
mapping. This implies that only a small part of all possible kinds
of mapping can slightly change the initial correlations in the
system. So, there is no point in coding every symbol (associating
every part of the phase space of the system with its binary code)
to analyze the correlation properties of the texts, as it is done,
for example, in Ref.~\cite{kokol}, but it is sufficient to use the
coarse-graining procedure.

One of the ways to get a correct insight into the nature of
correlations in a system consists in an ability of constructing a
mathematical object (for example, a correlated sequence of
symbols) possessing the same statistical properties as the initial
system. There exist many algorithms for generating long-range
correlated sequences: the inverse Fourier
transformation~\cite{czir}, the expansion-modification Li
method~\cite{li}, the Voss procedure of consequent random
additions~\cite{voss}, the correlated Levy walks~\cite{shl},
etc.~\cite{czir,her}. We believe that, among the above-mentioned
methods, using the many-step Markov chains is one of the most
important, because it offers a possibility to construct a random
sequence with definite correlation properties in the most natural
way. This was demonstrated in Ref.~\cite{uya}, where the concept
of additive Markov chain with the step-like memory function (which
allows the analytical treatment) was introduced. There exist some
dynamic systems (coarse-grained  sequences of Eukarya's DNA and
dictionaries) with the correlation properties that can be well
described by this model. The concept of additivity, primarily
introduced in paper~\cite{uyakm}, was later generalized for the
case of binary \emph{non-stationary} Markov chains~\cite{hod}.
Another generalization was based on consideration of Markov
sequences with a \emph{many-valued alphabet}~\cite{nar1, nar2}.

In the present work, we continue investigating into additive
Markov chains with more complex memory functions. An equation
connecting mutually-complementary characteristics of a random
sequence, i.e. the memory and correlation functions, is obtained.
Upon finding the memory function of the original random sequence
on the basis of the analysis of its statistical properties,
namely, its correlation function, we can build the corresponding
Markov chain, which possesses the same statistical properties as
the initial sequence. Effectiveness and robustness of the proposed
method is demonstrated by simple model examples. This method is
most essential for some applications, e.g., for the construction
of correlated sequence of elements which can be used to fabricate
the effective filters of electrical or optical signals,
Ref.~\cite{iku}. The suggested method allowed us to find memory
functions of concrete coarse-grained literary texts and to reveal
their universal power-law behavior at long distances.

Let us consider a homogeneous binary sequence of symbols,
$a_{i}=\{0,1\}$. To determine the $N$-\textit{step Markov chain}
we have to introduce the \emph{conditional probability}
$P(a_{i}\mid a_{i-N},a_{i-N+1},\dots ,a_{i-1})$ of occurring the
definite symbol $a_i$ (for example, $a_i =1$) after $N$-word
$T_{N,i}$, where $T_{N,i}$ stands for the sequence of symbols
$a_{i-N},a_{i-N+1},\dots ,a_{i-1}$. Thus, it is necessary to
define $2^{N}$ values of the $P$-function corresponding to each
possible configuration of the symbols in the $N$-word
$a_{i-N},a_{i-N+1},\dots ,a_{i-1}$. The value of $N$ is referred
to as the memory length of Markov chain.

Considering that we are going to deal with the sequences
possessing the memory length of order of $10^6$, we need to make
some simplification of the $P$-function. We suppose that it has
the \textit{additive} form,
\begin{equation}
P(a_{i}=1\mid T_{N,i})= \sum\limits_{k=1}^{N}f(a_{i-k},k),
\label{1}
\end{equation}
and corresponds to the additive influence of the previous symbols
upon the generated one. The value of $f(a_{i-k},k)$ is the
contribution of symbol $a_{i-k}$ to the conditional probability of
occurring the symbol unity at the $i$th site. The homogeneity of the
Markov chain is provided by the $i$-independence of conditional
probability Eq.~(\ref{1}).

Let us rewrite Eq.~(\ref{1}) in an equivalent form,
\begin{equation}
P(a_{i}=1\mid T_{N,i})=b+\sum\limits_{r=1}^{N} F(r)(a_{i-r}-b),
\label{2a}
\end{equation}
with
\begin{equation}\label{2c}
b=\frac{\sum\limits_{r=1}^{N}f(0,r)}{1-\sum\limits_{r=1}^{N}
F(r)}, \qquad F(r)=f(1,r)-f(0,r).
\end{equation}
The constant $b$ is the value of $a_{i}$ averaged over the whole
sequence, $b=\bar{a}$:
\begin{equation}\label{2e}
\bar{a}=\lim_{M\rightarrow
\infty}\frac{1}{2M+1}\sum_{i=-M}^{M}a_i.
\end{equation}
Indeed, according to the ergodicity of the Markov chain, $\bar{a}$
coincides with the value of $a_i$ averaged over the ensemble of
realizations of the Markov chain. So, we can write
\begin{equation}
\label{2b} \bar{a}=Pr(a_{i}=1)=\sum\limits_{T_{N,i}}P(a_{i}=1\mid
T_{N,i})Pr(T_{N,i}).
\end{equation}
Here $Pr(a_{i}=1)$ is the probability of occurring the symbol
$a_{i}$ equal to unity and $Pr(T_{N,i})$ is the probability of
occurring the definite word $T_{N,i}$ in the considering ensemble
of sequences. Substituting $P(a_{i}=1\mid T_{N,i})$ from
Eq.~(\ref{2a}) into Eq.~(\ref{2b}) and taking into account the
obvious relation $\sum\limits_{T_{N,i}}Pr(T_{N,i})=1$, one gets,
\begin{equation}\label{2d}
\bar{a}=b-b\sum\limits_{r=1}^{N}F(r) + \sum\limits_{r=1}^{N}F(r)
\sum\limits_{T_{N,i}}Pr(T_{N,i})a_{i-r}.
\end{equation}
The sum $\sum\limits_{T_{N,i}}Pr(T_{N,i})a_{i-r}$ does not depend
on the subscript $r$ and obviously coincides with $\bar{a}$. So,
we have $\bar{a}=b+(\bar{a}-b)\sum\limits_{r}F(r)$. From this
equation we conclude that $b=\bar{a}$. Thus, we can rewrite
Eq.~(\ref{2a}) as
\begin{equation}\label{2}
P(a_{i}=1 \mid T_{N,i})=\bar{a}+\sum\limits_{r=1}^{N}
F(r)(a_{i-r}-\bar{a}).
\end{equation}

We refer to $F(r)$ as the \emph{memory function} (MF). It describes
the strength of influence of previous symbol $a_{i-r}$ upon a
generated one, $a_{i}$. To the best of our knowledge, the concept of
memory function for many-step Markov chains was introduced in
Ref.~\cite{uyakm}. The function $P(. \mid .)$ contains the complete
information about correlation properties of the Markov chain.
Typically, the correlation function and other moments are employed
as the input characteristics for the description of the correlated
random sequences. However, the correlation function describes not
only the direct interconnection of the elements $a_i$ and $a_{i+r}$,
but also takes into account their indirect interaction via all other
intermediate elements. Our approach operates with the "origin"
characteristics of the system, specifically, with the memory
function. The correlation and memory functions are
mutual-complementary characteristics of a random sequence in the
following sense. The numerical analysis of a given random sequence
enables one to directly determine the correlation function rather
than the memory function. On the other hand, it is possible to
construct a random sequence using the memory function, but not the
correlation one. Therefore, we believe that the investigation of
memory function of the correlated systems will permit one to
disclose their intrinsic properties which provide the correlations
between the elements.

A dichotomic symbols in a Markov chain can be thought of as the
sequence of states of some particle, which participates in a
correlated Brownian motion. Every element of the sequence
corresponds to the instant change of particle's coordinate. Every
$L$-word (the sub-sequence of symbols of the length $L$ in the
sequence) can be regarded as one of the realizations of the
ensemble of correlated Brownian trajectories in the "temporal"
interval $L$. This point of view on the symbolic sequence makes it
possible to use the statistical methods for investigating the
dynamic systems.

We consider the distribution $W_{L}(k)$ of the words of definite
length $L$ by the number $k$ of unities in them,
$k_{i}(L)=\sum\limits_{l=1}^{L}a_{i+l}$, and the variance $D(L)$
of $k_{i}(L)$,
\begin{equation}\label{3}
D(L)=<(k-<k>)^2>,
\end{equation}
where the definition of average value of $g(k)$ is
$<g(k)>=\sum\limits_{k=0}^{L}g(k)W_{L}(k)$. It follows from
Eq.~(\ref{2}) that the positive MF values result in the
\emph{persistent} diffusion where previous displacements of the
Brownian particle in some direction provoke its consequent
displacement in the same direction. The negative values of the MF
correspond to the \emph{anti-persistent} diffusion where the
changes in the direction of motion are more probable. In terms of
the Ising model with long-range particles interactions that could
be naturally associated with the Markov chains, the positive
(negative) values of the MF correspond to the ferromagnetic
(anti-ferromagnetic) interaction of particles. The additive
form~(\ref{1}) of the conditional probability function corresponds
to the pair interaction and disregard of many-particles
interactions.

The memory function used in Refs.~\cite{uyakm,uya} was characterized
by the step-like behavior and defined by two parameters only: the
memory depth $N$ and the strength of symbol's correlations. Such a
memory function describes only one type of correlations in a given
system, the persistent or anti-persistent one, which results in the
super- or sub-linear dependence $D(L)$~\cite{rem}. Obviously, both
types of correlations can be observed at different scales in the
same system. Thus, one needs to use more complex memory functions
for detailed description of the systems with both type of
correlations. Besides, we have to find out a relation connecting the
mutually-complementary characteristics of random sequence, the
memory and correlation functions.

We suggest below two methods for finding the memory function
$F(r)$ of a random binary sequence with a known correlation
function. The first one is based on the minimization of a
"distance" $\emph{Dist}$ between the Markov chain generated by
means of a sought-for MF and the initial sequence of symbols. This
distance is determined by the formula,
\begin{equation}\label{optim1}
\emph{Dist}=\overline{(a_{i}-P(a_{i}=1 \mid T_{N,i}))^2} =
\lim_{M\rightarrow
\infty}\frac{1}{2M+1}\sum_{i=-M}^{M}(a_{i}-P(a_{i}=1 \mid
T_{N,i}))^2,
\end{equation}
with the conditional probability $P$ defined by Eq.~(\ref{2}).

Let us express distance~(\ref{optim1}) in terms of the correlation
function,
\begin{equation}\label{cor}
K(r)=\overline{a_{i}a_{i+r}}-\bar{a}^{2},\  \
K(0)=\bar{a}(1-\bar{a}),\  \ K(-r)=K(r).
\end{equation}
From Eqs.~(\ref{2}), (\ref{optim1}), one obtains
\[
\emph{Dist}=\sum\limits_{r,r'}\overline{(a_{i-r}-\bar{a})(a_{i-r'}
-\bar{a})}F(r)F(r')
-2\sum\limits_{r}\overline{(a_{i}-\bar{a})(a_{i-r}-\bar{a})}F(r)
+\overline{(a_{i}-\bar{a})^{2}}
\]
\begin{equation}
=\sum\limits_{r,r'}K(r-r')F(r)F(r')-2\sum\limits_{r}K(r)
F(r)+K(0). \label{optim3}
\end{equation}
The minimization equation,
\begin{equation}
\frac{\delta\emph{Dist}}{\delta
F(r)}=2\sum\limits_{r'}K(r-r')F(r')-2K(r)=0, \label{optim4}
\end{equation}
yields the relationship between the correlation and memory
functions,
\begin{equation} \label{main}
K(r)=\sum\limits_{r'=1}^{N}F(r')K(r-r'), \ \ \ \ r\geq 1.
\end{equation}
Equation~(\ref{main}) can also be derived by straightforward
calculation of the average $\overline{a_{i}a_{i+r}}$ in
Eq.~(\ref{cor}) using definition~(\ref{2}) of the memory function.
\protect\begin{figure}[h!]
\begin{centering}
\scalebox{0.8}[0.8]{\includegraphics{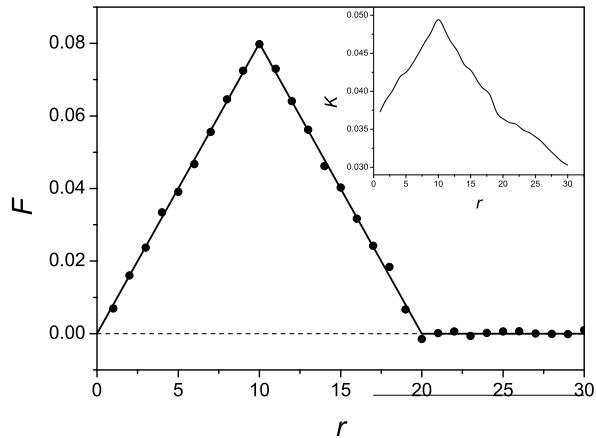}} \caption{The
initial memory function Eq.~(\ref{eqmf}) (solid line) and the
reconstructed one (dots) vs the distance $r$. In inset, the
correlation function $K(r)$ obtained by a numerical analysis of
the sequence constructed by means of the memory function
Eq.~(\ref{eqmf}).} \label{Fig1}
\end{centering}
\end{figure}

The second method resulting from the first one, establishes a
relationship between the memory function $F(r)$ and the variance
$D(L)$,
\begin{equation}
\label{MF} M(r,0)=\sum\limits_{r'=1}^{N}F(r')M(r,r'),
\end{equation}
\[
M(r,r')=D(r-r')-(D(-r')+r[D(-r'+1)-D(-r')]).
\]
It is a set of linear equations for $F(r)$ with coefficients
$M(r,r')$ determined by $D(r)$. The relations,
$K(r)=[D(r-1)-2D(r)+D(r+1)]/2$ obtained in Ref.~\cite{uyakm} and
$D(-r)=D(r)$ are used here.

\protect\begin{figure}[h!]
\begin{centering}
\scalebox{0.8}[0.8]{\includegraphics{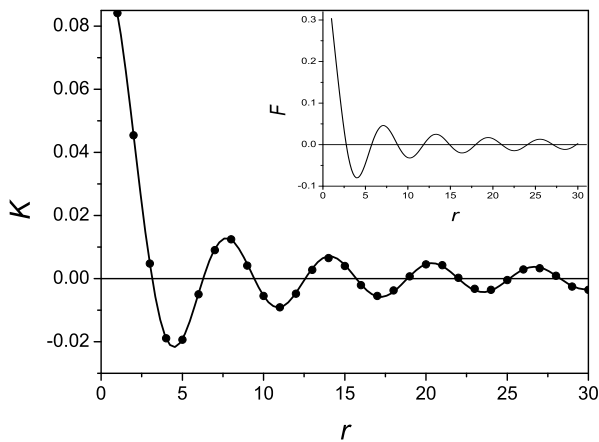}} \caption{The model
correlation function $K(r)$ described by Eq.~(\ref{K(r)}) (solid
line). The dots correspond to the reconstructed correlation
function. In inset, the memory function $F(r)$ obtained by
numerical solution of Eq.~(\ref{main}) with correlation function
Eq.~(\ref{K(r)}).} \label{Fig2}
\end{centering}
\end{figure}

Let us verify the robustness of our method by numerical
simulations. We consider a model "triangle" \emph{memory
function},
\begin{equation}
F(r)=0.008\cases {r,\;\qquad \;\;\ 1 \leq r < 10,  \cr 20-r,\;\;\
10 \leq r < 20 ,\cr 0, \;\;\;\;\;\;\;\;\;\;\ r\geq 20,}
\label{eqmf}
\end{equation}
presented in Fig.~\ref{Fig1} by solid line. Using Eq.~(\ref{2}),
we construct a random non-biased, $\bar{a}=1/2$, sequence of
symbols $\{0,1\}$. Then, with the aid of the constructed binary
sequence of the length $10^6$, we calculate numerically the
correlation function $K(r)$. The result of these calculations is
presented in inset Fig.~\ref{Fig1}. One can see that the
correlation function $K(r)$ mimics roughly the memory function
$F(r)$ over the region $1\leq r \leq 20$. In the region $r>20$,
the memory function is equal to zero but the correlation function
does not vanish~\cite{ref}. Then, using the obtained correlation
function $K(r)$, we solve numerically Eq.~(\ref{main}). The result
is shown in Fig.~\ref{Fig1} by dots. One can see a good agrement
of initial, Eq.~(\ref{eqmf}), and reconstructed memory functions
$F(r)$.

The main and very nontrivial result of our paper consists in the
ability to construct a binary sequence with an arbitrary
\emph{prescribed correlation function} by means of
Eq.~(\ref{main}). As an example, let us consider the model
correlation function,
\begin{equation}
K(r) = 0.1 \frac{\sin(r)}{r}, \label{K(r)}
\end{equation}
presented by the solid line in Fig.~\ref{Fig2}. We solve
Eq.~(\ref{main}) numerically to find the memory function $F(r)$
using this correlation function. The result is presented in inset
Fig.~\ref{Fig2}. Then we construct the binary Markov chain using
the obtained memory function $F(r)$. To check up a robustness of
the method, we calculate the correlation function $K(r)$ of the
constructed chain (the dots in Fig.~\ref{Fig2}) and compare it
with Eq.~(\ref{K(r)}). One can see an excellent agreement between
the initial and reconstructed correlation functions.

\protect\begin{figure}[h!]
\begin{centering}
\scalebox{0.8}[0.8]{\includegraphics{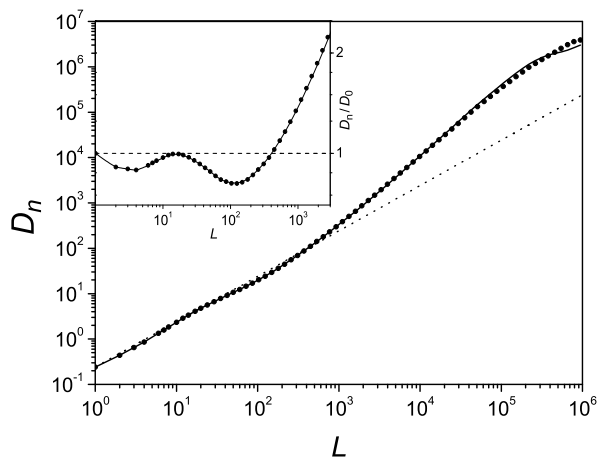}} \caption{The
normalized variance $D_n(L)$ for the coarse-grained text of Bible
(solid line) and for the sequence generated by means of the
reconstructed memory function $F(r)$ (dots). The dotted straight
line describes the non-biased non-correlated Brownian diffusion,
$D_{0}(L)=L/4$. The inset demonstrates the anti-persistent
dependence of ratio $D_n(L)/D_0(L)$ on $L$ at short distances.}
\label{Fig3}
\end{centering}
\end{figure}
Let us demonstrate the effectiveness of our concept of the
additive Markov chains when investigating the correlation
properties of coarse grained literary texts. First, we use the
coarse-graining procedure and map the letters of the text of
Bible~\cite{bibe} onto the symbols zero and unity (here, $(a-m)
\mapsto 0, (n-z) \mapsto 1$). Then we examine the correlation
properties of the constructed sequence and calculate numerically
the variance $D(L)$. The result of simulation of the normalized
variance $D_n(L)=D(L)/4 \bar{a}(1-\bar{a})$ is presented by the
solid line in Fig.~\ref{Fig3}. The dominator $4\bar{a}(1-\bar{a})$
in the equation for the normalized variance $D_n(L)$ is inserted
in order to take into account the inequality of the numbers of
zeros and unities in the coarse-grained literary texts. The
straight dotted line in this figure describes the variance
$D_{0}(L)=L/4$, which corresponds to the \textit{non-biased
non-correlated} Brownian diffusion. The deviation of the solid
line from the dotted one demonstrates the existence of
correlations in the text. It is clearly seen that the diffusion is
anti-persistent at small distances, $L\alt 300$, (see inset
Fig.~\ref{Fig3}) whereas it is persistent at long distances.

\begin{figure}[h!]
\begin{centering}
\scalebox{0.8}[0.8]{\includegraphics{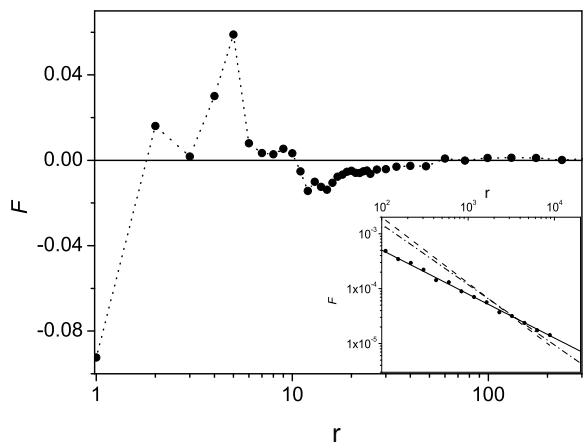}} \caption{The
memory function $F(r)$ for the coarse-grained text of Bible at
short distances. In inset, the power-law decreasing portions of
the $F(r)$ plots for several texts. The dots correspond to
"Pygmalion" by B. Shaw. The solid line corresponds to power-law
fitting of this function. The dash dotted and dashed lines
correspond to Bible in English and Russian, respectively.}
\label{Fig4}
\end{centering}
\end{figure}

The memory function $F(r)$ for the coarse-grained text of Bible at
$r<300$ obtained by numerical solution of Eq.~(\ref{MF}) is shown
in Fig.~\ref{Fig4}. At long distances, $r>300$, the memory
function can be nicely approximated by the power function
$F(r)=0.25r^{-1.1}$, which is presented by the dash-dotted line in
inset Fig.~\ref{Fig4}.

\protect

Note that the region $r \alt 40$ of negative anti-persistent
memory function provides much longer distances $L \sim 300$ of
anti-persistent behavior of the variance $D(L)$.

Our study reveals the existence of two characteristic regions with
different behavior of the memory function and, correspondingly, of
persistent and anti-persistent portions in the $D(L)$ dependence.
This appears to be a prominent feature of all texts written in any
language. The positive persistent portions of the memory functions
are given in inset Fig.~\ref{Fig4} for the coarse-grained English-
and Russian-worded texts of Bible (dash-dotted and dashed lines,
Refs.~\cite{bibe} and~\cite{bibr}, correspondingly). Besides, for
comparison, the memory function of the coarse-grained text of
"Pygmalion" by B. Shaw~\cite{pyg} is presented in the same inset
(dots), the power-law fitting is shown by solid line.

\begin{figure}[h!]
\begin{centering}
\scalebox{0.8}[0.8]{\includegraphics{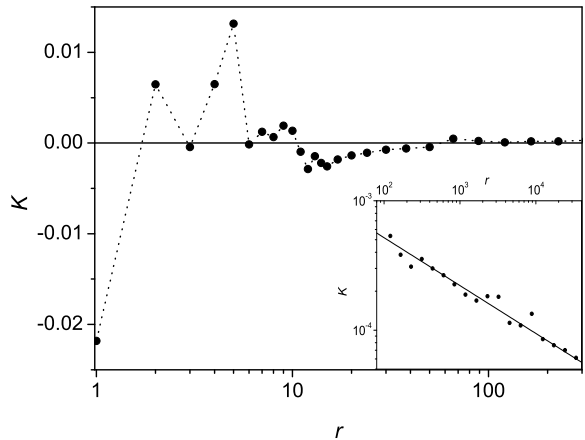}} \caption{The
correlation function $K(r)$ for the coarse-grained text of Bible
at short distances. In inset, the power-law decreasing portions of
the $K(r)$ plot for the same text. The solid line corresponds to
power-law fitting of this function.} \label{Fig5}
\end{centering}
\end{figure}

It is interesting to note that the memory function of any text
mimics the correlation function, as it was found for the model
example Eq.~(\ref{K(r)}). This fact is confirmed by
Fig.~\ref{Fig5} where the correlation function of the
coarse-grained text of Bible is shown. One can see that its
behavior at both short and long scales is similar to the memory
function presented in Fig.~\ref{Fig4}. However, the exponents in
the power-law approximations of $K(r)$ and $F(r)$ functions differ
essentially.

\protect

Thus, we have demonstrated the efficiency of description of the
symbolic sequences with long-range correlations in terms of the
memory function. An equation connecting the memory and correlation
functions of the system under study is obtained. This equation
allows reconstructing a memory function using a correlation
function of the system. Actually, the memory function appears to
be a suitable informative "visiting card" of any symbolic
stochastic process. The effectiveness and robustness of the
proposed method is demonstrated by simple model examples. Memory
functions for some concrete examples of the coarse-grained
literary texts are constructed and their power-law behavior at
long distances is revealed. Thus, we have shown the complexity of
organization of the literary texts in contrast to a previously
discussed simple power-law decrease of correlations~\cite{kant}.

Our theory describes not all statistical properties of the binary
symbolic sequences. For example, our consideration cannot reflect
the property of directivity of the texts since the theory is based
on the examination of the correlation function that is even by
definition. This property can be revealed only using the ternary
(or of the higher order) correlation functions. The
\emph{linguistic} aspects of the problem also require a regular
and systematic study.

We have examined the simplest examples or random sequences, the
dichotomic one. Nevertheless, our preliminary consideration
suggests that the presented theory can by generalized to the
arbitrary additive Markov process with a finite or infinite number
of states and with discrete or continuous "time". A study in this
direction is in progress.

The proposed approach can be used for the analysis of other
correlated systems in different fields of science.


\begin{references}
\bibitem{rem} Note that we here discuss the dependence of the
variance $D$ upon the length $L$ that describes the persistent (or
antipersistent) correlations in the words of different lengths
$L$. This length does not coincides with the memory length $N$, $L
\neq N$. The dependence $D(L)$ is completely different from the
dependence $D$ on the memory length $N$ discussed in
Ref.~\cite{nar1,nar2}.
\bibitem{ref} The existence of the "additional tail" in the correlation function
is in agreement with Ref.~\cite{uyakm} and corresponds to the well
known fact that the correlation length is always larger then the
region of memory function action.
\end{references}
\end{document}